\documentclass[main]{aa}
\usepackage{natbib}
\usepackage{graphicx}
\usepackage[english]{babel}
\usepackage{txfonts}
\usepackage{hyperref}
\usepackage{tabularx,booktabs}
\usepackage{multirow}
\usepackage{adjustbox}
\usepackage{amssymb}

\bibpunct{(}{)}{;}{a}{}{,}

\newcommand{\Rsun}{$R_{\sun}$}
\newcommand{\vmicro}{$\xi_{\rm t}$}
\newcommand{\kms}{km\,s$^{-1}$}
\newcommand{\te}{$T_{\rm eff}$}
\newcommand{\logg}{$\log{g}$}
\newcommand{\vsini}{$v\sin{i}$}

\begin{document}

\title
{Are the surface abundance structures stable in rapidly rotating Ap star 56~Ari?}

\author
{I.\,Potravnov\inst{1*},  N.\,Piskunov \inst{2} \and T.\,Ryabchikova\inst{1}} \institute
{Institute of Astronomy of RAS, Pyatnitskaya str., 48, 119017, Moscow, Russia\\
* e-mail: ilya.astro@gmail.com
\and Department of Physics and Astronomy, Division of Astronomy and Space Physics, Uppsala University, P.O. Box 516, 751 20, Uppsala, Sweden\\}

\date{}

\titlerunning{ }

\authorrunning{Potravnov et al.}

\abstract {The surface magnetic and abundance inhomogeneities in chemically peculiar Ap/Bp stars are coupled and responsible for their rotationally modulated variability. Within the framework of fossil field hypothesis these inhomogeneities are considered to be essentially stable over the Main Sequence (MS) timescale. However, a small group of Ap/Bp stars show rotational period changes, which are currently not well understood. We present results of Doppler Imaging (DI) of rapidly rotating Ap star 56 Ari for which changes in period were previously detected. Reconstruction of the surface distribution of silicon in 56 Ari reveals its complex spot pattern, which is responsible for the rotationally light variability and correlated with magnetic field modulation. Comparison of abundance maps obtained over the unprecedentedly long for such studies interval from 1986 to 2014 confirms stability and rigid rotation of the spot pattern. Thus, the period change in 56 Ari is not caused by rearrangement of the surface magnetic structures and/or atomic diffusion operating on short time scale. It is also unlikely to be explained by the visibility changes of the spots due to free-body precession of stellar rotational axis. In the end of the paper we briefly discuss possible alternative explanations of period variability.}
\maketitle

\section{Introduction}

The members of subgroup of chemically peculiar magnetic Ap/Bp stars constitute $\sim$10\% fraction of the early type A-B stars on the Main Sequence (MS). The atmospheres of these stars are characterised by the overabundance of many iron peak elements and heavier elements, some of them reaching up to several dex relative to the solar composition. Major fraction of Ap/Bp stars hosts globally organised magnetic fields, as strong as $\sim$10$^3$-10$^4$~G, although the origin of these fields remains a subject of debate. One of the prevailing hypotheses is the relic origin of these fields inherited by the stars from their parental molecular clouds \citep[see][for review]{Moss_2001,Braithwaite_2017}. In this paradigm, the magnetic field is frozen in the outer layers of the star and expected to be essentially stable over the MS lifetime of the star. 

The surface chemical composition of Ap/Bp stars is built up by the selective atomic diffusion \citep{Michaud_1970,Michaud_2015} in magnetised stellar atmosphere. This mechanism leads to the vertical chemical stratification of the elements depending on the interplay between gravity settlement and radiative acceleration for the atoms and ions. In the partially ionised plasma of a hot stellar atmosphere the magnetic field plays an important role in the diffusion process reducing the transport efficiency of the charged particles across the field lines. The accumulation of elements follows the surface geometry of the magnetic field \citep[e.g][]{Alecian_1981,Michaud_1981} resulting in the horizontal structures, i.e. chemical spots. Hence, the distribution of elements in Ap/Bp stellar atmospheres has the complex three-dimensional character, which modifies the thermal structure of the atmosphere and affects the emergent flux. Indeed, since the early history of investigations of Ap/Bp stars \citep{Belopolsky_1913,Guthnick_1914}, it is known that they exhibit a correlated photometric and line profile variability. Later it was supplemented by the magnetic variability proceeding in the same manner. Such a variability was explained by the oblique rotator model where the magnetic axis is inclined to the rotational one, and all light phenomena are caused by the rigid rotation of the spotted star \citep{Babcock_1949,Stibbs_1950}.

Because the built-up of vertical abundance stratification in the upper layers occurs on the timescale of few orders of magnitude shorter than changes in fossil magnetic fields due to stellar evolution or Ohmic diffusion \citep[e.g][]{Alecian_2006,Alecian_2011}, one can expect that the spotty structure and distribution should be essentially stable over the historical period of photometric and spectroscopic observations. Indeed, the extensive photometric studies revealed that for the vast majority of the Ap/Bp stars their spectro-photometric periods and light curves are precisely constant over several thousand rotational cycles \citep[e.g.][]{Preston_1971,Catalano_1998}. However, a small group of Ap/Bp stars was found to show the changing periods \citep{Musielok_1988,Pyper_1998,Adelman_2001,Mikulasek_2014,Adelman_2021}.

Based on a long-term photometric series, it was discovered that about of dozen of Ap/Bp stars exhibit a changes in the phases of their light curves extrema. The character of these changes is complex, while for some objects it can be described well by the model of linear period changes, few stars have shown marginal evidence for the variability in the shape of light curves \citep{Adelman_2021}. There is a unique case of CU Vir where the period change has a discrete character \citet{Pyper_1998,Pyper_2020}. It is still not clear whether these stars represent a homogeneous group, or what mechanism is responsible for the observed phase shifts. Given that both increase and (less frequently) decrease in the period were detected, as well as typical rates $\sim10^{-8}-10^{-9}$ d/cycle \citep{Mikulasek_2014,Adelman_2021}, these changes are hardly connected with the angular momentum evolution of the star as a whole. The more plausible explanation could be a variability in the spots distribution, hence challenging the assumption of the secular stability of surface magnetic and abundance structures. Alternatively, if the changes in the light curves are modulated with a secondary period, we may see a geometric effect of changing the visibility of the spots caused by precession of the magnetically distorted outer layers of a star \citep{Shore_1976}. 

A direct observational test that can shed some new light on the reasons for the puzzling period changes in Ap stars is the reconstruction of the horizontal abundance distributions over a time span essential for the manifestation of the above mentioned effect. The hot rapidly rotating star 56~Ari (sp:B9p Si) was one of the first for which a period change was suspected within the framework of the precession model \citep{Shore_1976}. The changes in the rotational period of the star $P\approx0.727$~d were  further detected by \citet{Musielok_1988} and \citet{Adelman_1993}. Whether there is a second period due to precession is not clear. The observed changes are described well by a linear spin down with the rate $\dot P\simeq 2-4.1$ s/100 yr \citep{Musielok_1988,Adelman_2001,Adelman_2021}. Preliminary results of Doppler Imaging (DI) of 56~Ari reported in \citet {Ryabchikova_2003} revealed the stability of the spot pattern over the 1986--2001 interval. Later, based on extensive photometric studies, \citet{Adelman_2021} proposed that if a precessional period exists in 56~Ari, its duration is at least $\sim$30 yr. In the present paper, we extend the analysis of the surface abundance distribution in 56~Ari for the comparable $\sim$30 yr time span in order to search for possible variations which could be responsible for the period changes observed for this star.

\section{Observational data}\label{Obs}

In our study, we collected phase-resolved spectroscopic time-series of 56~Ari over an interval of almost 30 years from 1986 to 2014. Our sample contains spectra observed in 1986-1987 at Lick Observatory \citep{Hatzes_1993} and provided by A. Hatzes, observations at Crimean Astrophysical Observatory obtained by V. Malanushenko in several seasons during 1996-2000  \citep{Adelman_2001}, 2004-2006 data obtained at Bohyunsan Optical Astronomy Observatory and provided by G. Valyavin \citep{Shulyak_2010}. We supplemented these data by the observations of 56~Ari obtained at Bernard Lyot Telescope with NARVAL spectrograph (PI: F. Lignieres, discussed in \citet{Shultz_2020}) and retrieved from the PolarBase archive \citep{Petit_2014}. In Table \ref{tbl1} we present the observational log, while some technical information on the instrumental setups is given below.

\begin{enumerate}
 \item Observations of the \ion{Si}{II} 6347 {\AA} line in 1986-87 were made using the coud{\'e} spectrograph of the 3-m Shane telescope at Lick Observatory equipped with the Texas Instruments 800$\times$800 CCD detector. The wavelength coverage for each observation  was 6332--6362 {\AA} at a spectral resolution of 0.13 {\AA} ($R\approx50000$). Slit losses due to seeing were minimised by using a Bowen-Walraven image slicer, which reformatted the light at the slit from a 3 arc-second hole to an equivalent of 0.67 arc-second slit. 
 \item Spectroscopic observations of 56~Ari were made in 1996--2001 with 2.6-m Shaijn telescope of the Crimean Astrophysical Observatory equipped by coud{\'e} spectrograph with the CCD detector. The spectrograms covered the 6325-6385 {\AA} region containing two strong \ion{Si}{II} 6347/6371 {\AA} lines. All spectra were obtained with spectral resolution of about 0.2 {\AA} (corresponding to resolving power $R\approx32000$). The signal-to-noise (S/N) ratios were in the range of 150-300.
 \item In 2004-2006 56~Ari was observed at Bohyunsan Optical Astronomy Observatory with 1.8-m telescope equipped by BOES fiber-fed echelle spectrograph with 3500-10000 {\AA} working range. 17 spectra with good phase coverage were obtained during this period. Observations were carried out in the medium-resolution mode, resulted in resolving power $R\approx30000$. The typical S/N$\approx$250-300 was achieved.
 \item 56~Ari was observed on 10 nights in September-October 2014 in spectropolarimetric mode of fibre-fed NARVAL echelle spectrograph at 2-m Bernard Lyot Telescope (at Pic du Midi). The spectrograms cover the 3700-10500 {\AA} range with resolving power $R\approx65000$. S/N$\approx$200-300 was achieved in Stokes $I$ spectra, which were used for further analysis.
\end{enumerate}

Data reduction for each spectrograph was described in the corresponding papers cited in the beginning of Section~\ref{Obs}.

All spectra except BOES and NARVAL were obtained in short spectral region around \ion{Si}{II} 6347/6371 {\AA} lines. Hence in our analysis we focused on this particular region. The spectra were carefully examined and only small corrections to the continuum level were introduced for better representation of the observations at close rotational phases. All spectra were wavelength shifted to compensate the stellar radial velocity $RV=+15.8$ \kms. The comparison revealed good stability of the wavelength solutions and $RV$'s over the entire observational period. A typical continuum level error does not exceed a few percent.\\
The NARVAL Stokes $I$ spectra were obtained from the PolarBase automatically processed and wavelength calibrated with the Libre-ESpRIT software \citep{Donati_1997}. In this instrumental configuration, \ion{Si}{II} 6347/6371 {\AA} lines fall on the edges of partly overlapping echelle orders. Comparison showed excellent agreement between the intensity and the shape of \ion{Si}{II} 6347 {\AA} line present in both spectral order. Therefore, the orders were merged and re-normalised to the continuum level using the low-order polynomial fit. The spectra were also RV-shifted to the stellar reference frame. The region of \ion{Si}{II} 6347/6371 {\AA} lines is blended with telluric water vapour features that were removed by dividing each observation of 56~Ari by an appropriately scaled spectrum of a rapidly rotating early-type star HR~1948 observed with the same NARVAL configuration which was retrieved from PolarBase.

\begin{figure}
	\includegraphics[width=1.0\linewidth]{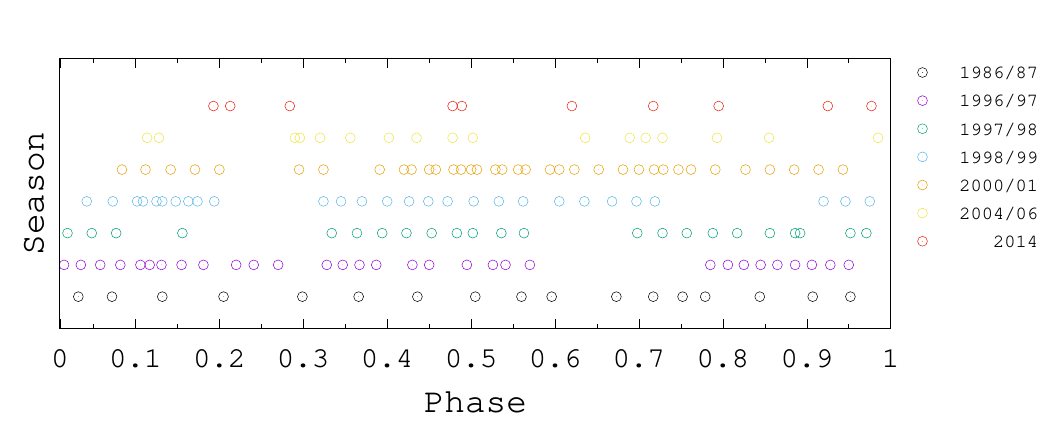}
    \caption{Phases distribution of 56~Ari observations.}
    \label{fig:1}
\end{figure}

All but BOES observations were obtained during the single observational seasons typically covered the period of autumn-winter visibility of 56~Ari. The BOES data were collected during several nights spread over 2004-2006 period. Generally our data has good phase coverage (Fig.~\ref{fig:1}), although gaps up to 0.2 in phase are presented in certain seasons, e.g. 1996/97, 1998/99 and in 2014. 

\section{Analysis and results}
\subsection{Stellar parameters and atmospheric model}

The most recent and accurate determination of the atmospheric parameters: effective temperature \te\, and surface gravity \logg\, for 56~Ari was made by \citet{Shulyak_2010} from the simultaneous fitting of the spectral energy distribution (SED) and hydrogen lines profiles in high resolution BOES spectra. They also determined abundances of few marker elements in 56~Ari atmosphere. We used this data summarised in Table~\ref{tab2} to calculate the model atmosphere of 56~Ari, which was further used for spectral synthesis. The calculations were made with the LLmodels code \citep{Shulyak_2004}, which builds plane-parallel stellar model atmospheres employing local thermodynamic equilibrium (LTE) assumption and line-by-line (LL) opacity treatment. This approach allows to account for the individual chemical composition and the line blanketing due to enhanced elemental abundances in the atmospheres of Ap/Bp stars. Following \citet{Shulyak_2010} we assume that mild magnetic field of 56~Ari with mean strength $<|B_Z|>\lesssim800$ G (see below) does not affect significantly the observed lines profiles and thus we further perform non-magnetic spectral synthesis. We adopted zero microturbulence \vmicro~ by analogy with other magnetic Ap/Bp stars, which is safe given the objectives of this study.

\begin{table}
\caption{Adopted parameters of 56~Ari}
\label{tab2}
\begin{tabular}{lcc}
\hline
Parameter & Value & Reference \\
\hline
\hline
\te & 12800$\pm$300 K & 1 \\
\logg  & 4.0$\pm$0.05 dex & 1 \\
\vmicro  & 0.0~\kms & 2 \\
\vsini   & 165$\pm$5~\kms & 2 \\
$i$      &$80^\circ\pm10$ & 2\\ 
$R/R_{\odot}$       & 2.38$\pm$0.2& 2\\
$\log (L/L_{\odot})$ & 2.14$\pm0.07$ & 2 \\
$M/M_{\odot}$       & 3.2$\pm0.3$ & 2,3 \\
$B_d$               & 2.7 kG & 4 \\
$\beta$               & 80-90$^\circ$ & 4,5 \\
\hline
\end{tabular}
\\ \\{\it Note. 1 - \citet{Shulyak_2010}; 2- this study; 3 - \citet{Kochukhov_2006}; 4 - \citet{Shultz_2020}; 5- \citet{Borra_1980}}
\end{table}    

The parameters also essential for the DI are the projected rotational velocity \vsini\, and the inclination angle $i$ of the rotational axis to the line of sight. The previous determinations of the rotational velocity in 56~Ari spectrum ranged from $\sim$75 to 160 \kms\, \citep{Abt_1995,Hatzes_1993}. \citet{Shultz_2020} found \vsini=153$\pm$10 \kms\, with the fitting of LSD profiles extracted from the NARVAL spectra. We also used the phase-averaged spectrum from the same 2014 NARVAL dataset to refine this parameter. Our estimation based on fitting the wings of \ion{Si}{II} 6347/6371\AA\, lines yielded \vsini=165$\pm$5 \kms\, in a satisfactory agreement with the Shultz' and Hatzes' values. \citet{Shulyak_2010} reported the radius $R=2.8\pm0.4$ \Rsun in their paper based on Hipparcos parallax of 56~Ari. Using the more up-to-date Gaia DR3 data \citep{GAIA_2023} we retrieved the parallax of 56~Ari $\pi=7.867\pm0.065$ mas, which can be converted to distance $D=127.1\pm1$ pc. Adopting this new distance we fitted the observational SED by the theoretical flux computed with LLmodels code for the adopted parameters and chemical composition of 56~Ari. The SED was constructed from ultraviolet to infrared using the spectrophotometry by \citet{Adelman_1983} supplemented by the data from the Two Micron All Sky Survey (2MASS, \citep{Skrutskie_2006}) and Wide-field Infrared Survey Explorer (WISE, \citep{Wright_2010}) catalogues. According to the statistical dust extinction map by \citet{Green_2019}, the interstellar reddening is almost negligible on the line of sight toward 56~Ari. Indeed, although the weak \ion{Na}{I} D interstellar lines are detectable in high resolution NARVAL spectrograms, their equivalent widths (EWs) calibrated against colour excess $E(B-V)$ with the \citet{Poznanski_2012} empirical relations provided visual extinction $A_V\lesssim0.05^m$ which can be neglected.

Finally, the SED was very well fitted by the theoretical flux scaled to the stellar radius $R=2.38\pm0.2$\Rsun\,. The mass of the star $M/M_{\odot}$ = 3.2$\pm0.3$ was determined from its position in the Hertzsprung-Russel diagram with the theoretical evolutionary tracks from the PARSEC model grid \citep{Bressan_2012}. Combining with the period at mid-observational epoch $\bar P=0.7279005$ d this yields within the errors the inclination angle $i\approx70-90^\circ$, implying that 56~Ari is observed almost equator-on. However, employing \vsini=153$\pm$10 \kms\, from \citet{Shultz_2020} we get a somewhat lower inclination $i=67^\circ$. We adopt value $i\approx80^\circ$, but in the later discussion we also consider the effect of a smaller inclination and projected rotational velocity on the DI results.  

The magnetic properties of the star have been investigated by \citet{Borra_1980,Shultz_2020}. Their magnetic $<|B_Z|>$ curve is shown in Fig.~\ref{fig:2}. These data separated by 34 yr are in reasonable agreement and show variations of longitudinal field of 56~Ari with $\approx$ 800~G amplitude. The magnetic curve fitting resulted in dipolar strength of magnetic field $B_d=2.7$~kG and large obliquity of a dipole with the angle between rotational and magnetic axes $\beta=80-90^\circ$. From the combined magnetic curve \citet{Shultz_2020} determined the period $P_{magn}=0.72776$, which is shorter than the rotational period determined from photometry. We also note that usage of the phases calculated with photometry-based variable period model results in $\Delta \phi\sim0.1$ shift between two magnetic curves.

\subsection{\ion{Si}{II} 6347/6371\AA\, variability in 56~Ari}

To examine the possible secular variations of silicon distribution over 56~Ari surface we first explore the EWs of \ion{Si}{II} 6347/6371\AA\,. The EWs were measured by direct integration of the lines profiles. As a result, phase curves of the EWs variations with rotational period were obtained for each of the available seasons. Using a constant period $P=0.727902$ d for phasing the data resulted in a noticeable phase shift indicative for a period increase, in agreement with previous results by \citet{Adelman_2001}. Hence we explore different period change models: the linearly increasing period with rates $\dot P=2$ s/100 yr \citep{Adelman_2001}, $\dot P=4.11$ s/100 yr \citep{Musielok_1988,Adelman_2021}, and model with two discrete period changes between 1963 and 2013 \citep{Adelman_2021}. Each of these variable period models results in much better phase agreement of the EW curves than using a constant period. Since EWs are generally less accurate than photometric data we accept the model with $\dot P=2$ s/100 yr which results in minimum phase dispersion of our data. Hence, the moments of observations $T$ were phased as $\varphi=\{(T-T_0)/P_{var}\}$ with variable period $P_{var}$ calculated as $P_{var}=P_0/(1+P_0\cdot S\cdot (T-T_0))$. Here, according to \citet{Adelman_2001}, $P_0=0.727883; T_0=2434322.354; S=-1.35\cdot10^{-9}$ is the retardation factor. 

\begin{figure}
	\includegraphics[width=1.0\linewidth]{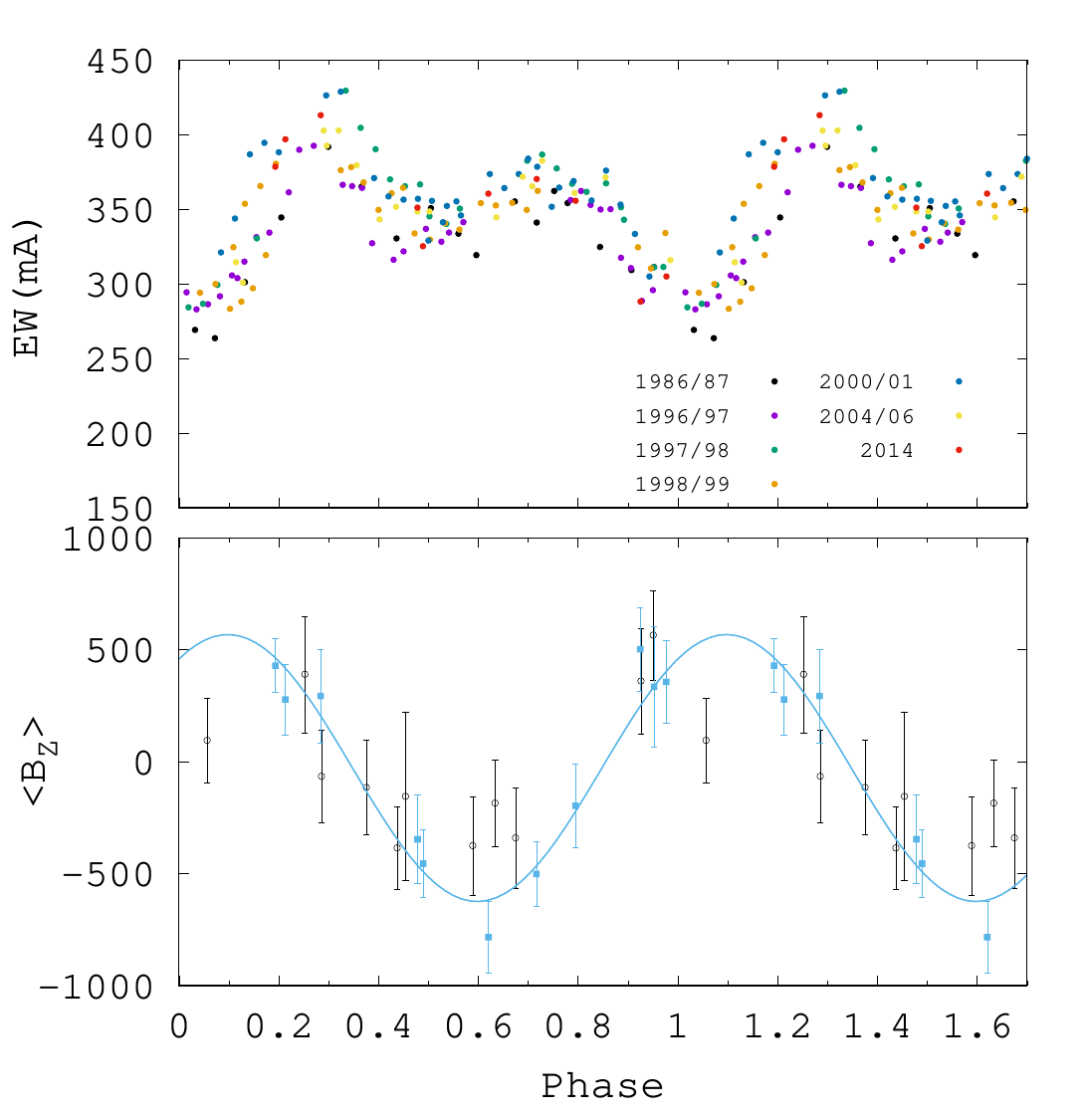}
    \caption{Upper panel: EWs of \ion{Si}{II} 6347\AA\, line phased with linear changing period model \citep{Adelman_2001}. Lower panel: variations of longitudinal magnetic field of 56~Ari according to \citet{Borra_1980} (black circles) and \citet{Shultz_2020} (blue squares) similarly phased to EW curve. The sinusoidal fit of the magnetic data set is shown by the blue curve. }
    \label{fig:2}
\end{figure}

The peak-to-peak EWs variations between different seasons reach $\sim$10\%, which is comparable with the estimated errors of continuum placement and EWs measurements. Indeed, for example, the adjustment of the continuum level by 7\% results in perfect match of EWs in 1986 and 2014 seasons. However, we did not use such artificial shifts in further handling of the data. The shape of EW curves (Fig.~\ref{fig:2}) remains essentially constant over the entire $\sim$30 yr time span. It is characterised by two maxima at phases $\varphi\approx0.3$ and $\varphi\approx0.7$, which also coincides with the maxima in photometric light curve. In Fig.~\ref{fig:2}, we compare the EWs variations and the magnetic field \citep{Borra_1980,Shultz_2020} observations phased with the adopted period change model. Given the large errors of magnetic measurements we tentatively conclude that the maxima of the EWs correspond to the null longitudinal field.  

\subsection{Doppler Imaging}

The surface distribution of silicon in 56~Ari was reconstructed with the help of INVERS11 code \citep{Piskunov_2002}. INVERS11 is a Doppler inversion code, which enables to fit the observed phase resolved variations of lines profiles by spectral synthesis with adjustable surface elemental distribution. The capability for synthesis of spectral line blends supports multi-element inversion. The code solves a non-magnetic radiative transfer equation computing the local line profile and then performs the disk integration accounting the Doppler shifts. As a typical ill-posed problem DI in INVERS11 is solved by introducing Tikhonov's regularisation to the map(s) of  elemental abundance(s). Fitting the observations is achieved by $\chi^2$-minimisation of the discrepancy between the calculated and the observed line profiles with the regularisation function scaled by the regularisation parameter $\Lambda$.

The \ion{Si}{} abundance maps at 7 epochs between 1986 and 2014 were obtained from fitting the spectroscopic timeseries of 56~Ari in the region of \ion{Si}{II} 6347/6371\AA\, lines. The LL-model atmosphere calculated with parameters from Table~\ref{tab2} and mean abundances from \citet{Shulyak_2010} were used for spectral synthesis. The atomic data for the transitions used in DI were critically compiled from the different sources with the help of VALD3 database \citep{Piskunov_1995,Ryabchikova_2015} and are summarised in Table~\ref{tab3}. We also adopted the projected rotational velocity \vsini=165 \kms\ and inclination angle $i=80^\circ$. However, for verification, we computed a set of maps also using a lower value \vsini=153 \kms\ \citep{Shultz_2020} and corresponding (for the adopted radius) inclination angle $i=68^\circ$. The results of the inversion appeared to be almost identical except for the visibility of hemispheres changed due to the inclination. This result is consistent with the error budget, which does not affect significantly the DI results \citep{Piskunov_1990} and estimated as $\sim\pm10^\circ$ in inclination and $\pm$5-10 \kms\, in velocity. 

Because \ion{Si}{II} 6347/6371\AA\, are blended by weak \ion{Fe}{II} and \ion{Mg}{II} lines we performed the multi-element fitting starting from the initial abundances from the adopted composition of 56~Ari. A series of calculations showed, however, the negligible contribution of the iron and magnesium to the observed variability of these blends. For the most realistic inversion the set of surface maps were calculated with the different values of the regularisation parameter $\Lambda$. The optimal solution was chosen for the $\Lambda$ yielding a minimum $\chi^2$ comparable with the S/N of our observations.

\begin{table}
\caption{List of elements and lines used for spectral synthesis and DI near \ion{Si}{II} 6347/6371\AA\,}
\label{tab3}
\tabcolsep=3pt
\begin{tabular}{lccccc}
\hline
Element & Line, \AA\, & $E_{low}$, eV & $\log gf$ & $\log \Gamma_4$ & Ref., $gf$ \\
\hline
\hline

\ion{Ni}{II} & 6341.7617 & 13.1774 &  0.716 & -5.150 & K03 \\
\ion{Fe}{II} & 6345.9282 & 11.0494 & -0.704 & -5.890 & RU \\
\ion{Ni}{II} & 6346.6614 & 12.9044 &  0.803 & -5.180 & K03 \\
\ion{Mg}{II} & 6346.7420 & 11.5690 &  0.020 & -3.500 & KP \\
\ion{Mg}{II} & 6346.7540 & 11.5690 & -1.280 & -3.500 & KP \\
\ion{Mg}{II} & 6346.9640 & 11.5691 & -0.140 & -3.500 & KP \\
\ion{Si}{II} & 6347.1087 & 8.1210  &  0.290 & -5.680 & Wilke \\
\ion{Si}{II} & 6347.1329 & 13.9351 & -1.421 & -2.810 & K14 \\
\ion{Si}{II} & 6347.1974 & 13.9352 & -1.267 & -2.810 & K14 \\
\ion{Fe}{II} & 6349.6029 & 11.0505 & -0.610 & -5.890 & RU \\
...\\
\ion{Fe}{II} & 6369.4590 & 2.8910  & -4.160 & -6.530 & VALD3 \\
\ion{Ni}{II} & 6371.2008 & 13.0802 & 0.815  & -5.120 & K03 \\
\ion{Si}{II} & 6371.3714 & 8.1210  & -0.020 & -5.680 & Wilke \\
\ion{Fe}{II} & 6371.7141 & 7.7083  & -1.692 & -5.720 & RU \\
\ion{Fe}{II} & 6372.4261 & 11.000  & -0.323 & -5.920 & RU \\
\ion{Fe}{II} & 6375.7964 & 10.9341 & -0.011 & -5.300 & RU \\

\hline
\end{tabular}
\\ \\{\it Note.} K03 = \citet{K03}, K14 = \citet{K14}, KP = \citet{KP}, Wilke = Wilke, R. 2003, Ph.D. Thesis, Heinrich-Heine-Universitat, Dusseldorf, RU = \citet{RU}, VALD3 = \citet{Ryabchikova_2015}
\end{table} 

\begin{figure*}
	\includegraphics[width=0.82\linewidth]{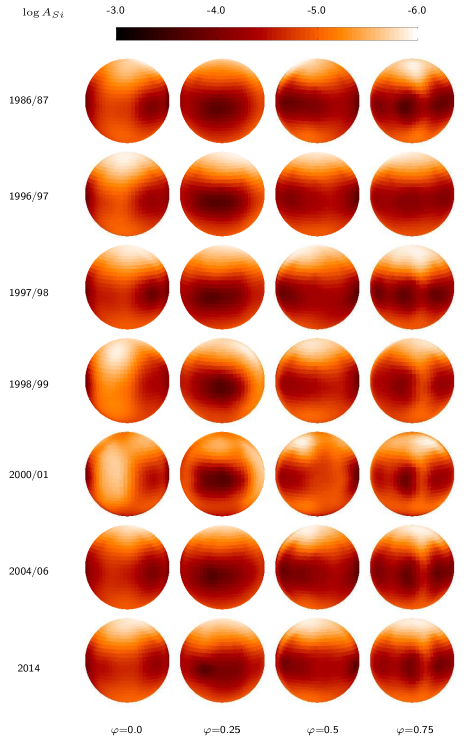}
    \caption{Distrubution of silicon over 56~Ari surface in 1986--2014. The spherical plots are presented for four equidistant phases. Abundance scale is given above the plots. Darker regions corresponds to higher \ion{Si}{} abundance.}
    \label{fig:3}
\end{figure*}

\begin{figure*}
	\includegraphics[width=1.0\linewidth]{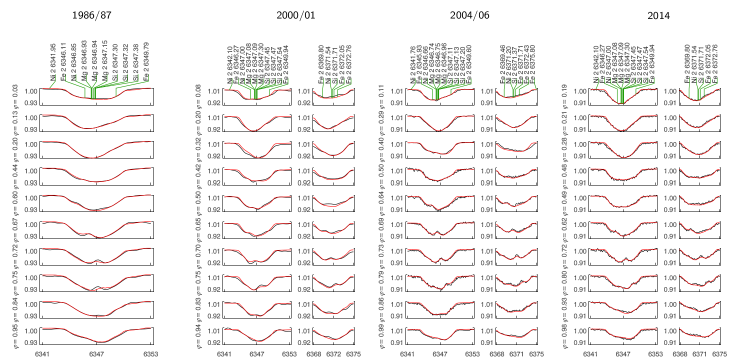}
    \caption{Examples of \ion{Si}{II} 6347/6371\AA\, profiles fitting in different seasons for different instrumental setups. Observational spectra are shown by black curves, synthesis - the red ones. Only 10 representative phases are shown in plot for convenience.}
    \label{fig:4}
\end{figure*}

The results of inversion procedure are shown in Fig.~\ref{fig:3}. The spherical plots are given for four equidistant rotational phases $\varphi=0-0.75$. Spatial resolution of the maps is of about $\Delta l\approx 5-11^\circ$ on the equator, depending on the spectral resolution of the dataset. One can see from the Fig.~\ref{fig:3} that 56~Ari possesses strongly inhomogeneous surface distribution of silicon, which is characterised by the two regions with the enhanced \ion{Si}{} abundance up to $\log A_{Si} = \log(N_{Si}/N_{tot})\approx$-3.0 dex separated by $\approx 180^\circ$ in longitude: the single large spot dominates at the phase $\varphi=0.25$ and very remarkable group of three spots at $\varphi=0.75$. The passages of these spots groups over visible hemisphere are responsible for the maximums on the EWs and light curves near corresponding phases. The spots possess some latitudinal elongation symmetrical relative to the equator, but this could be an artefact of DI because 56~Ari is seen almost equator-on making it difficult to differentiate between the southern and northern hemispheres. The northern polar region exhibits silicon depletion up to $\sim$1 dex with respect to the solar value ($\log A_{Si}^\odot$=-4.53). Another regions of the mild silicon underabundance in 56~Ari are vertical strips separating the spots groups and visible at phases $\varphi=0.0$ and $\varphi=0.5$. It is worth noting, that these regions of silicon depletion match with the extrema in magnetic curve. An example of profiles fitting is shown in Fig.~\ref{fig:4}. The most notable detail is the distortion of the profile by two transient absorption features in the phase range $\varphi\approx 0.6-0.8$, which corresponds to the passage of the spot triplet and reproduces reasonably well by the synthetic spectrum.

There is no evidence for any structural changes in the surface distribution of silicon in 56~Ari on the $\sim30$ yr interval between 1986 and 2014. A comparison of the surface maps in Fig.~\ref{fig:3} obtained at 7 epochs shows their excellent representation from season to season. During this period the number of silicon spots and their mutual position were preserved. Minor variations in the contrast of individual features as well as in abundance scale can be due to the phase coverage (see Fig.~\ref{fig:1}). Thus, an obvious consequence of these gaps is the degraded resolution near the 0.75 phase in 1996/1997 and 1998/99 seasons. Another consequence of both gaps and errors in continuum placement is the $\Delta A_{Si}\sim0.5$ dex variations of minimum in abundance scale from season to season. Hence the abundance scale in Fig.~\ref{fig:3} is normalised to the weighted mean value. We performed a simple numerical experiment computing the same set of maps but for a constant period $P=0.727902$ d \citep{Adelman_2001}. In this case, the longitudinal drift of the spots pattern responsible for the observed phases shift up to $\Delta\varphi\approx0.2$ between 1986 and 2014 is evident. Using the high-contrast vertical strip separating two silicon spots near phase $\varphi=$0.75 we are able to detect the longitudinal shift as small as $\sim$15 deg between 2000 and 2006 seasons.  This value is in agreement with (slightly exceeds) the resolution of our Doppler maps, hence with our technique we are able to detect the changes in the spots pattern of the same order of magnitude. We claim the absence of differential variation in the spot pattern in 56~Ari within this limit.

\section{Discussion}

56~Ari belongs to a small subgroup of magnetic Ap/Bp stars which show evidence for photometric period changes. Given yet not fully understood nature of this effect, different models are discussed. If the  model of linear period change is valid, then the rotational period of 56~Ari increases with a rate of $\dot P\simeq 2-4.1$ s/100 yr \citep{Musielok_1988,Adelman_2001,Adelman_2021}. Alternatively, there is a secondary period in variability of light curve shape with at least $\sim$30 years duration \citet{Adelman_2021}. These models imply completely different physical reasons. In the first case, the reason for period changes is either the spin down of the star or variations in the spot pattern from which the photometric period is deduced. In the latter case the reason for changes in the light curve shape could be the free-body precession of the star, which alters the visibility of the spots \citet{Shore_1976}. 

Using the DI technique we have reconstructed the surface distribution of silicon in 56~Ari and revealed the concentration of this element in the large scale spots with increased abundance. The results of the inversion are in very good agreement with the previous ones by \citet{Hatzes_1993,Ryabchikova_2003} that confirms reliability of the reconstructed spot pattern. Comparison of the spot distribution with the phase light curve (e.g. Fig.1 in \citet{Adelman_2001}) shows that the transit of regions with increased silicon abundance corresponds to the maxima in the light curve. In the temperature domain of Ap/Bp stars the \ion{Si}{II} is one of the most important opacity contributor \citep{Khan_2007} responsible for the modification of the temperature structure of atmosphere which caused the flux redistribution between UV and visible regions. In Ap stars, silicon spots are the major contributor to photometric variability \citep{Krtichka_2009,Krtichka_2012}. Hence, in the case of 56~Ari we can explicitly discuss the relation between the variations of the photometric period and distribution of silicon over the stellar surface.

We analysed the observations covering the  1986--2014 timeframe using DI techniques to investigate the secular behaviour of the spot pattern. Our main result is no differential changes in spots position over this $\sim30$ yr interval or $\sim10^4$ rotational cycles. One of the characteristic features of the star is the high obliquity of the magnetic axis with $\beta=80^\circ$ \citep{Borra_1980,Shultz_2020}. Thus, the magnetic axis lies almost in the plane of the rotational equator. Comparison of the magnetic phase curve with the surface maps, taking into account uncertainty of the former, reveals a silicon deficit in the regions of magnetic poles ($B_z$ extrema). The regions of increased silicon abundance thus are located in the plane of the magnetic equator although not forming a complete ring. However, for more confident conclusions about the correlation of \ion{Si}{} abundance distribution and magnetic field geometry the Zeeman-Doppler study of 56~Ari is highly desirable. 

Qualitatively these results are in agreement with diffusion calculations by \citet{Alecian_1981,Megessier_1984} predicting the silicon concentration near the magnetic equator at the ages of about $\sim10^7-10^8$ yr\footnote{We note that estimation of 56~Ari age is $\log t\approx7.95$ yr \citep{Kochukhov_2006}} and its subsequent migration toward the magnetic poles up to ages of order $\sim10^9$ yr. Thus, the horizontal diffusion in a presence of magnetic field has a timescale much longer than temporal coverage of our observations and timescale of period changes in 56~Ari. However, the timescale of vertical diffusion is 4-5 orders of magnitude shorter than for horizontal one \citep{Megessier_1984,Alecian_2009,Michaud_2015}, and hypothetically changes in the surface composition could be detected with the time-separated spectroscopic observations. Indeed, changes in the structure of chemical spots due to ongoing diffusion have been detected in a non-magnetic HgMn stars on the scale of a few years \citep{Kochukhov_2007} or even months \citep{Korhonen_2013}. In magnetic Ap/Bp stars the observable manifestation of diffusion effects requires either a stratification process that has not reached an equilibrium, or changes in the structure of the magnetic field. Based on our DI results we can rule out rapid changes in the number and structure of spots in 56~Ari due to diffusion on the timescale of the order $\sim10^1$ yr. This is also consistent with a recent observational result confirming the stability of the surface abundance distributions in the weakly magnetic star 45 Her on the 3 yr interval \citep{Kochukhov_2023}. Hence, rapidly operating diffusion is not the cause of the observed period changes in 56~Ari. 

If the silicon spot pattern in 56~Ari is stable on the $\sim$30 yr range, what can be physical reasons for the observed period changes? We will discuss some of the possibilities below in more details. 

\subsection{Evolutionary changes in period and magnetic braking}

First, we have to consider the possibility that the period change observed in 56~Ari is due to the angular momentum loss under the action of some braking mechanism. Rapidly-rotating early type stars often drive magnetised winds \citep{Babel_1997}, which can carry away a substantial fraction of angular momentum and lead to spin down of the host star \citep{Weber_1967,Mestel_1968}. The non-thermal radio emission observed in 56~Ari at 3 and 8.4 GHz \citep{Drake_2006} was attributed to the gyro-synchrotron radiation of relativistic electrons accelerated in the regions of wind-magnetosphere interaction \citep{Trigilio_2004,Leto_2021}. At the same time \citet{Shulyak_2010} suspected the weak distortion of the Balmer lines profiles in 56~Ari which could be produced by variable P Cyg feature arising in the stellar wind. However, \citet{Shultz_2020} reported no $H\alpha$ emission in 56~Ari and estimated the mass loss rate due to wind as $\log \dot M_W$=-11.7 $M_{\odot}$/yr. We adopt this value as the lower limit of the mass loss in the system. For rigidly rotating star we can write relation between period and angular momentum change $P/\dot P=J/\dot J\equiv \tau$, where $\tau$ is the characteristic braking timescale. The rate of the angular momentum $J\approx M_*\cdot R_*^2\cdot \Omega_*$ change can be expressed as $\dot J=2/3\cdot \Omega_*\cdot\dot M_W\cdot R_A^2$. Here $M_*,R_*,\Omega_*$ are the stellar mass, radius and angular velocity respectively, $R_A$ - Alfven radius. Substituting parameters of 56~Ari and adopting $R_A=45R_*$ \citep{Shultz_2020} we obtain $\tau_W\approx4\cdot10^8$ yr while the observed value of linear period increase is $P/\dot P\approx3.1\cdot10^6$ yr - two order of magnitude smaller. This difference probably can be explained by uncertainty in the mass loss rate and neglect its evolution. However, it is instructive to compare the observed deceleration time with the age of the star, which is $t\approx10^8$ yr. Thus, the observed value of $\dot P$ means that the mechanism responsible for the deceleration of 56~Ari was triggered at the time when the star was already on the MS. This is inconsistent with the assumption that in stars with luminosity not sufficient to accelerate massive radiative winds, the most efficient angular momentum loss may occur early in evolution by the accretion-driven wind at the pre-MS phase. Thus, it is unlikely that magnetic braking is the reason for the observed period increase in 56~Ari.     

\subsection{Possible free-body precession in 56~Ari}

It was pointed out that magnetic field can perturb the density distribution in the star and hence in oblique rotator (i.e. case of misalignment of rotational and magnetic axes) the free-body precession (Eulerian nutation) of the magnetically dominated outer envelope can arise about the axis different from a principal axis of inertia \citep{Spitzer_1957,Mestel_1972}. Later \citet{Shore_1976} proposed that in rapidly rotating Ap/Bp stars precession could produce an observable photometric and spectroscopic effects. Indeed, wobbling of rotational axis will lead to periodic modulation of the apparent latitudinal distribution of spots and their transit chords. Initially, in the paper cited above the precession period $p_{pr}$ for 56~Ari was estimated to be $\sim$5 yr. However, the subsequent analysis of the long-term photometric observations imposed a lower observational limit on the precession period $p_{pr}\gtrsim$30 yr \citep{Adelman_2021}. The surface maps of 56~Ari (Fig.~\ref{fig:3}) show almost perfect alignment of the silicon spots in 56~Ari along the equator. Comparison of the results for 1986--2014 seasons shows no significant changes in the latitudinal distribution of the spots within the error $\Delta l\approx10-15^\circ$. Thus, based on the Doppler inversion, we could not detect changes in the 56~Ari spot pattern due to precession with period $p_{pr}\lesssim$30 yr.
We also consider a well known result from the classical mechanics \citep[e.g.][]{Landau_1969}: for the spherically symmetric body with moments of inertia $I_1=I_2\neq I_3$) the precession period is $p_{pr}=P_{rot}\cdot I_1/[(I_1+I_3)\cdot \cos \chi]$, where $\chi$ is the angle between rotational axis and axis $I_3$. Obviously, when $\chi\rightarrow90^\circ$, then $p_{pr}\rightarrow\infty$. In the considered scenario of magnetic distortion for the 56~Ari $\chi\equiv\beta\approx80-90^\circ$. Thus, the orthogonality of the magnetic axis to the rotational one is strongly unfavourable to the arising of precession in this system or results in its very long period.

\subsection{Magneto-rotational effects}

Another possibility to explain the observed period increase is to abandon the principle that subsurface magnetic structures and spot pattern rotate rigidly with the bulk of the star. Such an assumption relaxes the difficulty in explaning rapid period changes by evolution of the angular momentum of the star as a whole. It was proposed by \citet{Stepien_1998} and further developed by \citet{Krticka_2017} that torsional oscillations driven by deviation from isorotation between magnetised envelope and stellar interiors \citep{Mestel_1987} could explain the period changes in CU Vir, if one interprets observational data as pointing to smooth periodical variations rather than discrete glitches.
On the other hand, it was supposed that subsurface magnetic fields of Ap/Bp stars might be shaped by Tayler instability \citep{Arlt_2011}. The non-axisymmetric mode of this kink-type instability of toroidal field in stellar interiors results in longitudinal propagation of global wave, which could shape the surface magnetic and abundance inhomogeneities. The remarkable feature is that instability pattern drifts in counter-rotational direction in the reference frame rotating with the star \citep{Rudiger_2010}. The counter-rotational migration of spots may lead to increase of photometrically-detected periods of Ap/Bp stars and, in particular, to be the reason for the phenomenon of super slowly rotating Ap stars \citep{Kitchatinov_2020}. Exploring the possibility that the period changes in 56~Ari are caused by the Tayler instability of its magnetic field will be the subject of forthcoming paper.

\section{Conclusions}

In the present work we examined the surface distribution of silicon in Ap star 56~Ari using DI technique over the interval from 1987 to 2014, which is unprecedentedly long compared to other related studies of Ap/Bp stars. We found that the distribution of \ion{Si}{} over the stellar surface is highly inhomogeneous and manifested as the group of equatorial spots. Regions of increased silicon abundance appear at the intersection of the magnetic and rotational equators, in agreement with the predictions of the diffusion theory. We found no differential changes in the number, structure, mutual position and abundance scale of spots on this $\sim$30 yr timespan. This result also implies stability of the configuration of the magnetic field frozen in the upper atmospheric layers during $\sim10^4$ rotational cycles.

On the other hand, 56~Ari is known to possesses the photometric period changes. Our result implies that secular variations in the \ion{Si}{} spot pattern are not the reason for these period changes. Within the accuracy limits of the DI method, we found no evidence for latitudinal variation of spots due to free-body precession that had been proposed in the literature as a potential reason for the changes in the light curve. Simple estimations suggest that angular momentum loss due to the magnetised stellar wind is also not a plausible reason. We speculate that magneto-rotational effects that preserve the rigid structure of the spot pattern but lead to their differential rotation with respect to deeper layers are a promising explanation. In particular, it is of interest to investigate the possible role of Tayler instability, which can lead to a longitudinal drift of magnetic and spot pattern in the counter-rotational direction and cause a changes in the photometric period.

\begin{acknowledgements} 
This research was funed by the grant of Russian Science Foundation \textnumero24-22-00237, https://rscf.ru/en/project/24-22-00237/.
\end{acknowledgements} 

\bibliographystyle{aa.bst}
\bibliography{references.bib}

\begin{appendix}
\section{Spectroscopic observations of 56~Ari}
\begin{table}
\caption{Journal of spectroscopic observations of 56~Ari \label{tbl1}}
\begin{tabular}{cc|cc|cc|cc}
\hline
HJD           & Phase & HJD          & Phase & HJD                     & Phase & HJD          & Phase \\
\hline
\multicolumn{2}{c}{1986-87}& \multicolumn{2}{c}{1997-98}  & \multicolumn{2}{c}{2000-01}    &  2453251.22843 &   0.708  \\                                     
2446719.92208 & 0.032 &      2450793.29102  &      0.019 &  2451976.19623 &   0.084        &  2453762.96817 &   0.728  \\                                     
2446722.86263 & 0.072 &      2450793.31229  &      0.048 &  2451976.21729 &   0.112        &  2453308.06812 &   0.793  \\                                     
2446719.99496 & 0.132 &      2450793.33353  &      0.077 &  2451976.23840 &   0.142        &  2453251.33591 &   0.855  \\
2446722.95909 & 0.205 &      2450783.20012  &      0.156 &  2451976.25947 &   0.171        &  2453760.97138 &   0.985  \\
2446814.74428 & 0.299 &      2450783.32930  &      0.334 &  2451976.28057 &   0.200        &\multicolumn{2}{c}{2014}   \\
2446809.69764 & 0.366 &      2450783.35151  &      0.364 &  2451979.26148 &   0.295        &  2456924.65618 &   0.193  \\
2446809.74838 & 0.436 &      2450783.37272  &      0.394 &  2451979.28261 &   0.324        &  2456957.42696 &   0.213  \\
2446750.83822 & 0.505 &      2450783.39389  &      0.423 &  2451918.18694 &   0.391        &  2456954.56719 &   0.284  \\
2446812.75031 & 0.560 &      2450783.41637  &      0.453 &  2451918.20818 &   0.420        &  2456955.43635 &   0.478  \\
2446750.90459 & 0.596 &      2450783.43756  &      0.483 &  2451861.43738 &   0.429        &  2456949.62120 &   0.489  \\
2446751.68841 & 0.673 &      2450783.45190  &      0.502 &  2451918.22943 &   0.450        &  2456958.45149 &   0.620  \\
2446662.91574 & 0.717 &      2450853.35587  &      0.536 &  2451861.45860 &   0.458        &  2456904.65544 &   0.717  \\
2446719.71780 & 0.752 &      2450797.32659  &      0.563 &  2451918.25048 &   0.479        &  2456950.57148 &   0.795  \\
2446813.63738 & 0.779 &      2450784.32260  &      0.698 &  2451861.48000 &   0.488        &  2456948.48213 &   0.925  \\
2446719.78510 & 0.844 &      2450784.34388  &      0.728 &  2451888.42158 &   0.500        &  2456913.57971 &   0.977  \\
2446751.85895 & 0.907 &      2450784.36502  &      0.757 &  2451918.27162 &   0.507        &                &          \\
2446719.86385 & 0.952 &      2450784.38794  &      0.788 &  2451888.44305 &   0.529        &                &          \\
\multicolumn{2}{c}{1996-97}& 2450784.40912  &      0.817 &  2451918.29270 &   0.537        &                &          \\
2450440.25090 & 0.015 &      2450784.43701  &      0.856 &  2451888.46239 &   0.556        &                &          \\
2450440.26537 & 0.035 &      2450784.45886  &      0.886 &  2451918.31377 &   0.565        &                &          \\
2450499.24264 & 0.058 &      2450787.37548  &      0.892 &  2451918.33482 &   0.594        &                &          \\
2450440.29944 & 0.082 &      2450793.24227  &      0.952 &  2451861.56516 &   0.605        &                &          \\
2450440.31720 & 0.106 &      2450793.25593  &      0.971 &  2451918.35589 &   0.623        &                &          \\
2450461.43420 & 0.117 &      \multicolumn{2}{c}{1998-99} &  2451918.37696 &   0.652        &                &          \\
2450440.33498 & 0.131 &      2451213.28616   &     0.005 &  2451918.39804 &   0.681        &                &          \\
2450440.35278 & 0.155 &      2451160.22414   &     0.109 &  2451924.23526 &   0.700        &                &          \\
2450462.20903 & 0.181 &      2451240.30569   &     0.124 &  2451919.15257 &   0.718        &                &          \\
2450462.23762 & 0.220 &      2451240.32349   &     0.149 &  2451924.25639 &   0.729        &                &          \\
2450462.25235 & 0.241 &      2451187.19564   &     0.162 &  2451919.17375 &   0.747        &                &          \\
2450403.31300 & 0.270 &      2451240.34173   &     0.174 &  2451924.28003 &   0.762        &                &          \\
2450467.41102 & 0.328 &      2451187.21933   &     0.195 &  2451924.30109 &   0.791        &                &          \\
2450467.42548 & 0.347 &      2451241.17923   &     0.324 &  2451924.32729 &   0.827        &                &          \\
2450467.43970 & 0.367 &      2451241.19389   &     0.345 &  2451924.34849 &   0.856        &                &          \\
2450467.45400 & 0.387 &      2451241.21197   &     0.369 &  2451924.36956 &   0.885        &                &          \\
2450409.25299 & 0.430 &      2451241.23413   &     0.400 &  2451924.39066 &   0.914        &                &          \\
2450409.26758 & 0.450 &      2451222.34386   &     0.449 &  2451924.41174 &   0.943        &                &          \\
2450428.22600 & 0.495 &      2451161.23952   &     0.504 &  \multicolumn{2}{c}{2004-06}    &                &           \\
2450428.24811 & 0.526 &      2451161.26094   &     0.533 &  2453309.02950 &   0.114        &                &           \\
2450428.25893 & 0.541 &      2451161.28216   &     0.563 &  2453306.12820 &   0.128        &                &           \\
2450428.28060 & 0.570 &      2451161.31322   &     0.605 &  2453309.16206 &   0.290        &                &           \\
2450501.22824 & 0.785 &      2451161.33465   &     0.635 &  2453759.00984 &   0.296        &                &           \\
2450501.24315 & 0.806 &      2451161.35919   &     0.668 &  2453250.21911 &   0.320        &                &           \\
2450501.25735 & 0.825 &      2451161.38049   &     0.698 &  2453306.29422 &   0.356        &                &           \\
2450501.27150 & 0.845 &      2451161.39587   &     0.719 &  2453762.00264 &   0.402        &                &           \\
2450429.22295 & 0.865 &      2451213.22445   &     0.920 &  2453250.30179 &   0.435        &                &           \\
2450501.30160 & 0.886 &      2451213.24290   &     0.946 &  2453671.06895 &   0.478        &                &           \\
2450501.31579 & 0.906 &      2451213.26418   &     0.975 &  2453671.08601 &   0.502        &                &           \\
2450501.33214 & 0.928 &                      &           &  2453666.08805 &   0.636        &                &           \\
2450501.34826 & 0.950 &                      &           &  2453760.02840 &   0.689        &                &           \\

\hline	  
\end{tabular}
\\ \\{Note. Rotational phases are given according to ephemeris for linearly changing period from \citet{Adelman_2001}}
\end{table} 
\end{appendix}

\end{document}